
\hsize=6.0truein\vsize=8.5truein\voffset=0.25truein
\hoffset=0.1875truein
\tolerance=1000\hyphenpenalty=500
\def\monthintext{\ifcase\month\or January\or February\or
   March\or April\or May\or June\or July\or August\or
   September\or October\or November\or December\fi}

\def\monthdayandyear{\monthintext\space\number\day, \number\year}

\font\tenrm=cmr10 scaled \magstep1   \font\tenbf=cmbx10 scaled \magstep1
\font\sevenrm=cmr7 scaled \magstep1  
\font\fiverm=cmr5 scaled \magstep1   
\font\teni=cmmi10 scaled \magstep1   \font\tensy=cmsy10 scaled \magstep1
\font\seveni=cmmi7 scaled \magstep1  \font\sevensy=cmsy7 scaled \magstep1
\font\fivei=cmmi5 scaled \magstep1   \font\fivesy=cmsy5 scaled \magstep1

\font\tentt=cmtt10 scaled \magstep1
\font\tenit=cmti10 scaled \magstep1
\font\tensl=cmsl10 scaled \magstep1
\def\twelvepoint{\def\rm{\fam0\tenrm}
   \textfont0=\tenrm \scriptfont0=\sevenrm \scriptscriptfont0=\fiverm
   \textfont1=\teni  \scriptfont1=\seveni  \scriptscriptfont1=\fivei
   \textfont2=\tensy \scriptfont2=\sevensy \scriptscriptfont2=\fivesy
   \textfont\itfam=\tenit \def\it{\fam\itfam\tenit}
   \textfont\ttfam=\tentt \def\tt{\fam\ttfam\tentt}
   \textfont\bffam=\tenbf \def\bf{\fam\bffam\tenbf}
   \textfont\slfam=\tensl \def\sl{\fam\slfam\tensl} \rm
   \hfuzz=1pt\vfuzz=1pt
   \setbox\strutbox=\hbox{\vrule height 10.2pt depth 4.2pt width 0pt}
   \parindent=24pt\parskip=1.2pt plus 1.2pt
   \topskip=12pt\maxdepth=4.8pt\jot=3.6pt
   \normalbaselineskip=13.5pt\normallineskip=1.2pt
   \normallineskiplimit=0pt\normalbaselines
   \abovedisplayskip=13pt plus 3.6pt minus 5.8pt
   \belowdisplayskip=13pt plus 3.6pt minus 5.8pt
   \abovedisplayshortskip=-1.4pt plus 3.6pt
   \belowdisplayshortskip=13pt plus 3.6pt minus 3.6pt
   \topskip=12pt \splittopskip=12pt
   \scriptspace=0.6pt\nulldelimiterspace=1.44pt\delimitershortfall=6pt
   \thinmuskip=3.6mu\medmuskip=3.6mu plus 1.2mu minus 1.2mu
   \thickmuskip=4mu plus 2mu minus 1mu
   \smallskipamount=3.6pt plus 1.2pt minus 1.2pt
   \medskipamount=7.2pt plus 2.4pt minus 2.4pt
   \bigskipamount=14.4pt plus 4.8pt minus 4.8pt}
\twelvepoint

\def\letter{\parindent=0pt\parskip=\medskipamount\def\endmode{}
   \def\longindent{\parindent=0.0truein\obeylines\parskip=0pt}
   \def\letterhead{\null
      \begingroup\longindent
      \def\endmode{\medskip\medskip\endgroup}}
   \def\date{\endmode\begingroup\longindent
      \def\endmode{\medskip\medskip\endgroup}\monthdayandyear}
   \def\address{\endmode\begingroup
      \parindent=0pt\obeylines\parskip=0pt
      \def\endmode{\medskip\medskip\endgroup}}
   \def\salutation{\endmode\begingroup
      \parindent=0pt\obeylines\parskip=0pt\def\endmode{\medskip\endgroup}}
   \def\body{\endmode\begingroup\parskip=\medskipamount
      \def\endmode{\medskip\medskip\endgroup}}
   \def\closing{\endmode\begingroup\longindent
      \def\endmode{\endgroup}}
   \def\signed{\endmode\begingroup\longindent\vskip0.8truein
      \def\endmode{\endgroup}}
   \def\endofletter{\endmode \ifnum\pageno=1 \nopagenumbers\fi
      \vfil\vfil\eject}}

\def\preprint#1{ 
   \def\draft{\finishtitlepage{PRELIMINARY DRAFT:
\monthdayandyear}\writelabels%
      \headline={\sevenrm PRELIMINARY DRAFT: \monthdayandyear\hfil}}
   \def\date##1{\finishtitlepage{##1}}
   \font\titlerm=cmr10 scaled \magstep3
   \font\titlerms=cmr10 scaled \magstep1 
   \font\titlei=cmmi10 scaled \magstep3  
   \font\titleis=cmmi10 scaled \magstep1 
   \font\titlesy=cmsy10 scaled \magstep3 	
   \font\titlesys=cmsy10 scaled \magstep1  
   \font\titleit=cmti10 scaled \magstep3	
   \skewchar\titlei='177 \skewchar\titleis='177 
   \skewchar\titlesy='60 \skewchar\titlesys='60 
   \def\titlefont{\def\rm{\fam0\titlerm}
      \textfont0=\titlerm \scriptfont0=\titlerms 
      \textfont1=\titlei  \scriptfont1=\titleis  
      \textfont2=\titlesy \scriptfont2=\titlesys 
      \textfont\itfam=\titleit \def\it{\fam\itfam\titleit} \rm}
   \def\title##1{\vskip 0.9in plus 0.4in\centerline{\titlefont ##1}}
   \def\authorline##1{\vskip 0.9in plus 0.4in\centerline{\bf ##1}
      \vskip 0.12in plus 0.02in}
   \def\author##1##2##3{\vskip 0.9in plus 0.4in
      \centerline{{\bf ##1}\myfoot{##2}{##3}}\vskip 0.12in plus 0.02in}
   \def\addressline##1{\centerline{##1}}
   \def\abstract{\vskip 0.7in plus 0.35in\centerline{\bf Abstract}\smallskip}
   \def\finishtitlepage##1{\vskip 0.8in plus 0.4in
      \leftline{##1}\supereject\endgroup}
   \baselineskip=19pt plus 0.2pt minus 0.2pt \pageno=0
   \begingroup\nopagenumbers\parindent=0pt\baselineskip=13.5pt\rightline{#1}}

\def\nolabels{\def\eqnlabel##1{}\def\eqlabel##1{}\def\figlabel##1{}%
   \def\reflabel##1{}}
\def\writelabels{\def\eqnlabel##1{%
   {\escapechar=` \hfill\rlap{\hskip.11in\string##1}}}%
   \def\eqlabel##1{{\escapechar=` \rlap{\hskip.11in\string##1}}}%
   \def\figlabel##1{\noexpand\llap{\string\string\string##1\hskip.66in}}%
   \def\reflabel##1{\noexpand\llap{\string\string\string##1\hskip.37in}}}
\nolabels
\global\newcount\secno \global\secno=0
\global\newcount\meqno \global\meqno=1
\def\newsec#1{\global\advance\secno by1
   \xdef\secsym{\the\secno.}
   \global\meqno=1\bigbreak\medskip
   \noindent{\bf\the\secno. #1}\par\nobreak\smallskip\nobreak\noindent}
\xdef\secsym{}
\def\appendix#1#2{\global\meqno=1\xdef\secsym{\hbox{#1.}}\bigbreak\medskip
\noindent{\bf Appendix #1. #2}\par\nobreak\smallskip\nobreak\noindent}
\def\acknowledgements{\bigbreak\medskip\centerline{\bf
   Acknowledgements}\par\nobreak\smallskip\nobreak\noindent}
\def\eqnn#1{\xdef #1{(\secsym\the\meqno)}%
	\global\advance\meqno by1\eqnlabel#1}
\def\eqna#1{\xdef #1##1{\hbox{$(\secsym\the\meqno##1)$}}%
	\global\advance\meqno by1\eqnlabel{#1$\{\}$}}
\def\eqn#1#2{\xdef #1{(\secsym\the\meqno)}\global\advance\meqno by1%
	$$#2\eqno#1\eqlabel#1$$}
\def\myfoot#1#2{{\baselineskip=13.5pt plus 0.3pt\footnote{#1}{#2}}}
\global\newcount\ftno \global\ftno=1
\def\foot#1{{\baselineskip=13.5pt plus 0.3pt\footnote{$^{\the\ftno}$}{#1}}%
	\global\advance\ftno by1}
\global\newcount\refno \global\refno=1
\newwrite\rfile
\def\ref{[\the\refno]\nref}
\def\nref#1{\xdef#1{[\the\refno]}\ifnum\refno=1\immediate
   \openout\rfile=\jobname.aux\fi\global\advance\refno by1\chardef\wfile=\rfile
   \immediate\write\rfile{\noexpand\item{#1\ }\reflabel{#1}\pctsign}\findarg}
\def\findarg#1#{\begingroup\obeylines\newlinechar=`\^^M\passarg}
   {\obeylines\gdef\passarg#1{\writeline\relax #1^^M\hbox{}^^M}%
   \gdef\writeline#1^^M{\expandafter\toks0\expandafter{\striprelax #1}%
   \edef\next{\the\toks0}\ifx\next\null\let\next=\endgroup\else\ifx\next\empty%
\else\immediate\write\wfile{\the\toks0}\fi\let\next=\writeline\fi\next\relax}}
   {\catcode`\%=12\xdef\pctsign{
\def\striprelax#1{}
\def\semi{;\hfil\break}
\def\addref#1{\immediate\write\rfile{\noexpand\item{}#1}} 
\def\listrefs{\vfill\eject\immediate\closeout\rfile
   \centerline{{\bf References}}\medskip{\frenchspacing%
   \catcode`\@=11\escapechar=` %
   \baselineskip=15pt plus 0.2pt minus 0.2pt
   \input \jobname.aux\vfill\eject}\nonfrenchspacing}
\def\startrefs#1{\immediate\openout\rfile=refs.tmp\refno=#1}
\global\newcount\figno \global\figno=1
\newwrite\ffile
\def\fig{\the\figno\nfig}
\def\nfig#1{\xdef#1{\the\figno}\ifnum\figno=1\immediate
   \openout\ffile=\jobname.fig\fi\global\advance\figno by1\chardef\wfile=\ffile
   \immediate\write\ffile{\medskip\noexpand\item{Fig.\ #1:\ }%
   \figlabel{#1}\pctsign}\findarg}
\def\listfigs{\vfill\eject\immediate\closeout\ffile{\parindent48pt
   \baselineskip16.8pt\centerline{{\bf Figure Captions}}\medskip
   \escapechar=` \input \jobname.fig\vfill\eject}}

\def\noblackbox{\overfullrule=0pt}
\def\inv{^{\raise.18ex\hbox{${\scriptscriptstyle -}$}\kern-.06em 1}}
\def\dup{^{\vphantom{1}}}
\def\Dsl{\,\raise.18ex\hbox{/}\mkern-16.2mu D} 
\def\dsl{\raise.18ex\hbox{/}\kern-.58em\partial}
\def\slash#1{\raise.18ex\hbox{/}\kern-.68em #1}
\def\boxeqn#1{\vcenter{\vbox{\hrule\hbox{\vrule\kern3.6pt\vbox{\kern3.6pt
   \hbox{${\displaystyle #1}$}\kern3.6pt}\kern3.6pt\vrule}\hrule}}}
\def\mbox#1#2{\vcenter{\hrule \hbox{\vrule height#2.4in
   \kern#1.2in \vrule} \hrule}}  
\def\bar{\overline}\def\psibar{\bar\psi}
\def\e#1{{\rm e}^{\textstyle#1}}
\def\del{\partial}
\def\curly#1{{\hbox{{$\cal #1$}}}}
\def\curlyL{\hbox{{$\cal L$}}}
\def\vev#1{\langle #1 \rangle}
\def\lform{\hbox{$\sqcup$}\llap{\hbox{$\sqcap$}}}
\def\darr#1{\raise1.8ex\hbox{$\leftrightarrow$}\mkern-19.8mu #1}
\def\half{{\textstyle{1\over2}}} 
\def\roughly#1{\ \lower1.5ex\hbox{$\sim$}\mkern-22.8mu #1\,}
\def\MSbar{$\bar{{\rm MS}}$}
\hyphenation{di-men-sion di-men-sion-al di-men-sion-al-ly}
\def\ket#1{\vert #1 \rangle}\def\bra#1{\langle #1 \vert}
\def\Half{{1\over2}}
\def\d#1#2{d\mskip 1.5mu^{#1}\mkern-2mu{#2}\,}
\def\mev{\mathop{\rm Me\kern-0.1em V}\nolimits}
\def\gev{\mathop{\rm Ge\kern-0.1em V}\nolimits}
\def\alphaS{{\alpha_S}}\def\alphapi{{\alpha_S\over4\pi}}
\def\npb#1,#2,#3 {Nucl.\ Phys.\ {\bf B#1} (19#2) #3}
\def\prd#1,#2,#3 {Phys.\ Rev.\ D {\bf #1} (19#2) #3}
\def\plb#1,#2,#3 {\ifnum#1>170 Phys.\ Lett.\ B {\bf #1} (19#2) #3\else
 Phys.\ Lett.\ {\bf #1B} (19#2) #3\fi}
\def\sjnp#1,#2,#3 {Sov.\ J.\ Nucl.\ Phys.\ {\bf #1} (19#2) #3}
\def\yadfiz#1,#2,#3 {Yad.\ Fiz.\ {\bf #1} (19#2) #3}


\font\fivesl=cmsl8 scaled 750 
\def\kappac{\kappa_{\fivesl c}}
\def\kappaconeloop{\kappa_{c\thinspace{\fivesl one-loop}}}
\def\uoneloop{u_{0\thinspace{\fivesl one-loop}} }

\def\third{\textstyle{1\over3}} 
\def\curlyO{\curly O}
\def\onezero{({\bf 1}\thinspace\thinspace{\bf 0})}
\def\bdagger{b^\dagger}
\def\fsubB{$f_B$}
\def\zerohat{{\bf \hat 0}}
\def\muhat{\hat \mu}
\noblackbox
\preprint{hep-lat/9401035}
\rightline{UdeM-LPN-TH-93-183}
\rightline{UCLA/93/TEP/42}
\vskip -0.2in
\title{Tadpole-Improved Perturbation Theory }
\smallskip
\centerline{ \titlefont for Heavy-Light Lattice Operators }
\vskip 0.35in plus 0.10in
\centerline{\bf Oscar F. Hern\'andez}
\vskip 0.1in plus 0.02in
\addressline{ Laboratoire de physique nucl\'eaire}
\addressline{ Universit\'e de Montr\'eal C.P. 6128}
\addressline{ Montr\'eal, Qu\'e., Canada H3C 3J7}
\addressline{ 
oscarh@lps.umontreal.ca}
\vskip 0.25in plus 0.05in
\centerline{\bf Brian R. Hill}
\vskip 0.1in plus 0.02in
\addressline{Department of Physics}
\addressline{University of California}
\addressline{Los Angeles, CA~~90024}
\addressline{ bhill@physics.ucla.edu}
\vskip -0.1in
\abstract
Lattice calculations of matrix elements involving heavy-light quark
bilinears are of interest in calculating a variety of properties of
$B$ and $D$ mesons, including decay constants and mixing parameters.
A large source of uncertainty in the determination of these properties
has been uncertainty in the normalization of the lattice-regularized
operators that appear in the matrix elements.  Tadpole-improved
perturbation theory, as formulated by Lepage and Mackenzie, promises
to reduce these uncertainties below the ten per cent level at
one-loop.  In this paper we study this proposal as it applies to
lattice-regularized heavy-light operators.  We consider both the
commonly used zero-distance bilinear and the distance-one point-split
operator.  A self-contained section on the application of these
results is included.  The calculation reduces the
value of $f_B$ obtained from lattice calculations using the heavy
quark effective theory.
\date{January 1994}
\newsec{Introduction}%
For a heavy quark such as the $b$ quark, a variety of\nref\lm{G.~P.~Lepage and
P.~B.~Mackenzie, Phys. Rev. D {\bf 48} (1993) 2250.}
approaches are available for lattice Monte Carlo calculations,
including non-relativistic QCD (NRQCD)~\ref\NRQCD{
W. E. Caswell and G. P. Lepage, \plb167,86,437\semi
{G.~P.~Lepage and B. A. Thacker},~ in {\sl Field Theory on the Lattice,}
edited by A.~Billoire {\it et al.}, Nucl. Phys. B (Proc. Suppl.)
{\bf 4} (1988) 199\semi
{B.~A.~Thacker and G.~P.~Lepage}, Phys. Rev. D {\bf 43} (1991) 196\semi
G.~P.~Lepage, {\it et al}, Phys. Rev. D {\bf 46} (1992) 4052.},
the heavy quark effective theory (HQET)~\ref\HQET{H. D. Politzer
and M. B. Wise, \plb208,88,504\semi
{E.~Eichten and B. Hill},~ \plb234,90,511\semi
B.~Grinstein,~ Nucl. Phys. B {\bf 339}, 253 (1990)\semi
H.~Georgi, \plb240,90,447.},
extrapolation from Wilson fermion results~\ref\extrapolate{``A
Lattice Computation of the Decay Constants of B and D Mesons,''
C.~W.~Bernard, J.~N.~Labrenz, and A.~Soni, UW--PT--93--06,
to appear in Phys. Rev. D.}, and
formulations which interpolate continuously between these
actions~\ref\interpolate{A.~X.~El-Khadra,  A.~S.~Kronfeld, and
P.~B.~Mackenzie, ``Massive Quarks in Lattice QCD,''
FERMILAB--PUB--93/195--T, in preparation.}.
Whatever approach is used, the relationship of the lattice operators
to the operators coming from the continuum electroweak theory must
be calculated in order to make use of the lattice results.  While
these short-distance strong interaction corrections are in principle
perturbatively calculable, in practice, the one-loop corrections are
sometimes so large as to be of questionable reliability
\ref\ehtwo{
E.~Eichten and B.~Hill, \plb240,90,193\semi
Ph.~Boucaud,~ C.~L.~Lin and O.~P\`ene, Phys.\ Rev.\ D {\bf 40},
1529 (1989)\semi
{\it ibid}, {\bf 41}, 3541(E) (1990).}\nref\hhone{O.~F.~Hern\'andez
and B.~R.~Hill, \plb237,90,95.
}--\nref\hhtwo{O.~F.~Hern\'andez and B.~R.~Hill, \plb280,92,91.}%
\ref\fhh{J.~M.~Flynn, O.~F.~Hern\'andez and B.~R.~Hill, Phys. Rev. D {\bf 43}
(1991) 3709.}.  It is possible that
two-loop corrections are so large as to make existing calculations
very inaccurate at realistically attainable values
of $\beta=6/g_0^2$.

It is disturbing that these large corrections exist and they
seem to point to a deficiency in our understanding of lattice perturbation
theory.  Recently Lepage and Mackenzie~\lm\ have suggested
a reorganization of perturbation theory, the root of which is a
nonperturbatively determined renormalization of the basic
operators in the lattice action.  These redefinitions remedy the problem
of large renormalizations arising from lattice tadpole graphs.

Another problem with lattice perturbation theory is that if one
uses $\beta$ to determine the perturbative coupling, $\alpha_{\fivesl lat}$,
one-loop perturbative corrections in
quantities such as the mass renormalization for Wilson fermions are
consistently underestimated.  These perturbation theory problems
are due to the fact that $\alpha_{\fivesl lat}$ is a poor choice of
expansion parameter.  For example at an inverse lattice spacing of 2~GeV,
Lepage and Mackenzie's tadpole-improved expansion parameter is
$\alpha_V = 0.16$,
which is twice as large as $\alpha_{\fivesl lat} = 0.08$.  The reason that
$\alpha_{\fivesl lat}$ is so different from $\alpha_V$ is that
$\alpha_{\fivesl lat}$ receives large all orders corrections from the
tadpoles present in lattice QCD.  One is led to the new expansion
parameter after one redefines the basic fields in the lattice action
and follows the tadpole improvement program referred to in the previous
paragraph.  Lepage and Mackenzie argue that the best way to arrive
at $\alpha_V$ is from a non-perturbative lattice determination
of a perturbatively calculable quantity, such as
the gauge field plaquette expectation value.
After the reorganization of perturbation theory and the determination
of a suitable expansion parameter,
Lepage and Mackenzie reanalyze
several lattice quantities
noted
for their poor agreement with
perturbative calculations, and they find that the results of their
program are in good agreement with the Monte Carlo results.

The scope of the present paper is a
study of the Lepage-Mackenzie prescription as it
applies to the calculation of quantities determined using the heavy quark
effective action, particularly $f_B$.  In the following section, we will
review the effect of tadpole improvement on the heavy and light quark
actions.  To keep the explanation simple, discussions of one-loop perturbative
corrections are postponed to Section~3.  Then in Section~4, we
determine the coupling and the scale for which the coupling should be
evaluated as it appears in the expressions of Section~3.  In Section~5,
we complete the lattice-to-continuum matching.

Our main results are in Tables~4a and~4b of Section~6.
This concluding section has been made as
self-contained as possible, and readers who just want to know how to
apply our results can
turn directly to it.
There we incorporate the results of additional continuum calculations
of the next-to-leading running and matching from
the continuum effective theory at the lattice scale to the
full theory at the scale~$m_{\fivesl b}$~\ref\nextToLeading{X.~Ji and
M.~J.~Musolf,
\plb257,91,409\semi
{D.~J.~Broadhurst and A.~G.~Grozin},~ \plb267,91,105\semi
{C.~A.~Dominguez and N.~Paver},~ \plb293,92,197.}\
which is not otherwise discussed here.  More detail on
the application of those results can be found in
Ref.~\ref\bigpaper{A.~Duncan, E.~Eichten, J.~Flynn, B.~Hill, and H.~Thacker,
Fermilab--PUB--93/377--T, in preparation.}.
The results are
tabulated in a form in which they can be easily used to obtain
physical predictions from lattice results.
Throughout the paper, it is assumed that the light quark action is the
standard Wilson fermion action.  Both staggered fermion and naive fermion
results correspond to the case $r=0$~\hhone.
However the conventional
normalization of the Wilson fermion field is singular as $r\rightarrow0$,
and a different convention is used in that case.  This is discussed in the
conclusions.

\newsec{Tadpole Improvement of the Heavy and Light Quark Actions}%
Using tadpole improvement of the Wilson action for quarks on the
lattice as a guide, one can perform tadpole
improvement of the heavy quark action, and this has been done
by Bernard in Ref.~\ref\bernard{C.~Bernard, ``Heavy-Light and
Light-Light Weak Matrix Elements on the Lattice,'' plenary talk
at Lattice '93, Dallas, TX, USA, hep-lat/9312086.}.
Instead of discretizing
\eqn\contaction{
S=\int d^4x \thinspace b^\dagger \left(i \partial_0 + g A_0\right) b
}
as
\eqn\usual{
S=i a^3 \sum\nolimits_{n}
b^\dagger(n)
\left(
b(n)-\vphantom{1\over u_0}U_0(n{-}\zerohat)^\dagger b(n{-}\zerohat)
\right)\!.
}
the analog of the Lepage and Mackenzie prescription for the Wilson action is
\eqn\lmrec{
S_{\fivesl tadpole-improved}=i a^3 \sum\nolimits_{n}
b^\dagger(n)
\left(
b(n)-{1\over u_0}U_0(n{-}\zerohat)^\dagger b(n{-}\zerohat)
\right)\ ,
}
where $u_0$ is defined as
\eqn\uodef{
u_0 \equiv \vev{\third {\rm Tr} U_{\fivesl plaq}}^{1/4}\ .
}
The combination $U_\mu(x)/u_0$ more closely
corresponds to the continuum field $(1+iagA_\mu(x))$, than does
$U_\mu(x)$ itself.

Since Monte-Carlo calculations have been performed with the former
of the two actions in Eqns.~\usual\ and \lmrec,
we now derive the
relationship between matrix elements determined with the two actions.
Consider the matrix element
\eqn\martrix{
\vev{J^{0\dagger}_5(n_0,{\bf 0}) J^0_5(0)}={1\over a^3} G_B(n_0)\ ,
}
where
\eqn\jofive{
J^0_5(n) \equiv b^\dagger(n) ({\bf 1 \ 0} )\gamma^0 \gamma_5 q(n)\ .
}
The two-by-four matrix $({\bf 1 \ 0})$ is the upper two rows of the
projection operator $(1 + \gamma_0)/2$.  In general our notation for
heavy quark fields and heavy-light currents follows that of Ref.~\hhtwo.
As one finds by performing the $b$ and $b^\dagger$ integrations
in the lattice-regularized functional integral in a fixed gauge field
configuration $\{U_\mu(n)\}$, the heavy quark propagator is just the product
\eqn\productOfLinks{
U_0(n_0-1,{\bf 0})^\dagger ... U_0(0,{\bf 0})^\dagger.
}
With the tadpole-improved action, there is an additional factor of $1/u_0$ for
each gauge field link in the product.  Thus
\eqn\relat{
[G_B(n_0)]_{\fivesl tadpole-improved} = {G_B(n_0) \over u_0^{n_0} }
}

The $B$ meson decay constant $f_B$ is usually extracted from numerical
simulations by fitting $G_B(n_0)$ to
\eqn\fit{
{(f_B m_B)^2\over 2 m_B} \exp[-Cn_0 a]
}
As discussed in Ref.~\ehtwo, for perturbative corrections,
this necessitates using the reduced value of the heavy quark
wave function renormalization, the constant part of which at one-loop is,
$c=4.53$.

{}From Eqns.~\relat\ and~\fit, we see that the tadpole improvement procedure
has no effect on the fitted value of $f_B$.  Its only effect is the change
\eqn\lmfit{
C\rightarrow C+{\ln u_0\over a},
}
that is, a linearly divergent mass renormalization.

Interestingly, if one uses the other fitting procedure discussed in
Ref.~\ehtwo, the application of the Lepage-Mackenzie prescription leads one to
introduce factors of $u_0$ which led to a final result identical to
the above.  Thus the reduction of $c$ from 24.48 to 4.53 discussed in
Ref.~\ehtwo\ is the same reduction caused by the reorganization of the
tadpole contributions~\bernard.

We now consider the point-split operator.  A variety of operators with the same
continuum limit as in Eq.~\jofive\ can be constructed.
Consider the set of distance-one bilinears which respects the remnants of the
$O(3)$ rotational group that is present in the lattice heavy quark effective
theory.  As discussed in Ref.~\hhtwo, the only operator in this set that
one needs to consider is
\def\ihat{\hat \imath}
\eqn\spac{
{J_{\fivesl ps}^{~0}}_5(n) \equiv
{1\over6}\sum_i\left[
\bdagger(n{+}\ihat)U_i(n)^\dagger({\bf 1~0})\gamma^0\gamma_5 q(n)+
\bdagger(n{-}\ihat)U_i(n{-}\ihat) ({\bf 1~0})\gamma^0\gamma_5 q(n)
\right].}
The index $i$ runs only over the three spatial directions.
The sum of six terms has been chosen to respect the remnants of the $O(3)$
rotational group.  The tadpole improvement procedure for this operator is
to multiply it by a single factor of $1/u_0$ since it contains a single
gauge field link.

So far in this section, we have seen that tadpole
improvement does not affect the extraction of $f_B$ as it
is generally done in lattice Monte Carlo calculations using the
zero-distance bilinear, and that the effect of tadpole improvement on
the distance-one bilinear receives a contribution of $1/u_0$.
However we must still take into account
the effect of tadpole improvement of the light quark action, and this will
involve some additional factors.

As conventionally defined in lattice Monte Carlo calculations
the lattice operator $J^0_5$ involved in
calculating $f_B$ is renormalized by a factor
$\sqrt{2\kappac}Z$, where $\kappac$ is the critical value of the
hopping parameter for the light quarks.  The tree level value of
$\kappac=1/(8r)$.  What
Lepage and Mackenzie are advocating is a reorganization of perturbation
theory so that the $\kappac$ factors are included in $\tilde Z$ and the
renormalizing factor becomes
$\tilde Z/(2\sqrt{r})$.~\myfoot{$^\dagger$}{This is actually a slight
generalization of the Lepage-Mackenzie prescription to the case $r\ne1.$}
To the extent that
the non-perturbatively calculated value of $\kappac$ agrees with its
perturbatively calculated counterpart, there is no difference in these two
prescriptions.
In the next section we will see how to proceed with this procedure at one-loop
order.

\newsec{ Tadpole Improvement of Heavy-Light Operators at One-Loop Order }%
This section will present in parallel
the analysis for the zero-distance and
point-split lattice discretizations of $J^0_5$.
To summarize the previous section, we saw that
rather than multiplying the zero-distance operator
by $\sqrt{2\kappac}Z$, it
should be multiplied by $\tilde Z/(2\sqrt{r})$,
and that rather than multiplying the
distance-one operator by $\sqrt{2\kappac}Z_{\fivesl ps}$,
it should be multiplied
by ${\tilde Z}_{\fivesl ps}/(2 u_0\sqrt{r})$.
In this section we will calculate and explain the relationship
between $Z$ and $\tilde Z$ (and
$Z_{\fivesl ps}$ and ${\tilde Z}_{\fivesl ps}$) at one-loop order.

Suppose that calculations of $Z$ and $Z_{\fivesl ps}$ have been carried out to
one-loop order and that the result is
\eqn\ZandZps{
\eqalign{
Z          & = 1 + {\alpha_S \over 3 \pi} [
\int {d^4 q\over \pi^2} g(q) + {3\over2}\ln(\mu a)^2] \ ,
\cr
Z_{\fivesl ps} & = 1 + {\alpha_S \over 3 \pi} [
\int {d^4 q\over \pi^2} g_{\fivesl ps}(q) + {3\over2}\ln(\mu a)^2] \ .
\cr}}
Suppose further that a one-loop calculation of $8\kappac$ has been performed,
and that the result is
\eqn\oneLoopKappaC{
{1\over 8r\kappaconeloop} = 1 - {\alpha_{\fivesl S}\over 3\pi }
\int {d^4q \over \pi^2} h(q)
}
(we will reserve the unadorned symbols $\kappac$ and $u_0$ for the
non-perturbative lattice Monte Carlo values, and qualify
them with the subscript ``one-loop'' when referring to their
perturbative values).
Finally, suppose that a one-loop calculation of $u_0$ gives,
\eqn\oneLoopPlaquette{
\uoneloop = 1 + {\alpha_S\over 3\pi }
\int {d^4q \over \pi^2} j(q)\ .
}
Then any quantity obtained from a lattice calculation, for which
short-distance corrections have been correctly included at one-loop order
can be multiplied by
\eqn\fancyWaysOfWritingOne{
{\left(8 r  \kappaconeloop\over 8 r \kappac\right)^p}
{\left(\uoneloop\over u_0\right)^s}
}
where $p$ and $s$ are any small numbers, and the result remains correct
at one-loop order.

Without some kind of additional consideration there
is no reason to prefer one of these one-loop results over another.
In terms of the various quantities defined above,
the tadpole improvement requires that
for the $Z$~factor of the distance-zero operator, $p=1/2$ and $s=0$.
In other words the relationship between $\tilde Z$ and $Z$ is
\eqn\zrelat{
\tilde Z = \sqrt{8r\kappaconeloop} Z \ .
}
Thus if an expression
for $\tilde Z$ analogous to the expression for $Z$
in Eqn.~\ZandZps\ were defined, we would have
\eqn\gtwiddleofq{\tilde g(q) = g(q) + h(q)/2\ .}
Similarly, for the point-split operator,
$p=1/2$ and $s=1$, implying
\eqn\zrelatps{
\tilde Z_{\fivesl ps}
= \uoneloop\sqrt{8r\kappaconeloop}Z_{\fivesl ps}
}
and
\eqn\gtwiddlepsofq{\tilde g_{\fivesl ps}(q) = g_{\fivesl ps}(q) + h(q)/2 +
j(q)\ .}
For clarity we emphasize that the relationship between $Z$ and
$\tilde Z$ is not $\tilde Z = \sqrt{8r\kappac} Z$ and $\tilde Z_{\fivesl ps} =
u_0 \sqrt{8r\kappac} Z_{\fivesl ps}$.
In that case, tadpole improvement would be without content, since
we would be renormalizing by the exact same factor in both cases.

The functions $g(q)$, $h(q)$, $g_{\fivesl ps}(q)$, and $j(q)$ have
already been calculated, and we now assemble the results.
For various frequently appearing functions of $q$, we will use the
notation $\Delta_i$ as defined in Ref.~\ref\berdrapson{C.~Bernard, A.~Soni,
and T.~ Draper, Phys. Rev. D {\bf 36} (1987) 3224.}. (Note that $\Delta_6$
defined in Ref.~\fhh\ conflicts with $\Delta_6$ defined in Ref.~\berdrapson.)
We also define
$\Delta_i^{(3)}$ to be the same as $\Delta_i$ but with $q_0=0$.
{}From Refs.~\ehtwo\ and~\hhone, we have
\eqn\gofq{
\eqalign{
  g(q)  &=-{1\over 4\Delta_1\Delta_2} + {\theta(1-q^2)\over q^4}
         + {1 - 4 r^2\over 16 \Delta_2}
	 - {r\over 4\Delta_2^{(3)}} + {1 \over 16\pi^2} \cr
         &\phantom{=} -{1\over2}
         (-{1\over8\pi^2}+{2\over 16 \Delta_1^2}-{2\theta(1-q^2)\over q^4}-
         {1\over32\Delta_1})
         \cr
         &\phantom{=} - {1\over2}
         ({1\over 32\pi^2}+ {1\over 8\Delta_1}+{\theta(1-q^2)\over q^4}
         - {\Delta_4 + \Delta_5\over16\Delta_1^2\Delta_2}
         + {r^2(2-\Delta_1)\over4\Delta_2}) \ .\cr
}}
The second and third groups are $-c/2$ ($c$ is the reduced heavy
quark wave function renormalization discussed in Section~2) and $-(f-F)/2$
($f-F$ is the constant part of the wave function renormalization
for a Wilson fermion), respectively.

The critical hopping parameter $\kappac$ is
related to the critical bare mass by the expression
\eqn\masscritical{
{1\over8r\kappac}= 1 + {m_c a\over4r} \ .
}
Thus for massless Wilson
quarks only the linearly divergent part of the correction to $m_c$
survives in the continuum limit.
At one-loop, in terms of $h(q)$ defined in \oneLoopKappaC, the result
is~\berdrapson\ref\guido{G.~Martinelli, \plb141,84,395.},
\eqn\hq{
h(q)={1\over4}\bigl({1\over 2 \Delta_1} - {1\over 4 \Delta_1
\Delta_2}[2\Delta_1\Delta_6 +
\Delta_4] \bigr) \ .
}
The first term comes from the tadpole graph and
has the larger value after integration.
\goodbreak

{}From Ref.~\hhtwo, we find
\eqn\gpsofq{
  g_{\fivesl ps}(q) = g(q) +
         {4\Delta_1-\Delta_1^2 + 2\Delta_4 + 12 r^2 \Delta_1^2 \over 24
\Delta_1 \Delta_2} - {r\Delta_1^{(3)}\over 6\Delta_2^{(3)}}\ .
}
where $g(q)$ is defined in Eq.~\gofq.
Finally,
\eqn\jofq{
  j(q)= -{1\over16}\ .
}
The integral of $j(q)$ is determined by the coefficient of the first term
in Eq.~(4.1).

The constant part of the one-loop corrections for any
value of the Wilson mass term coefficient $r$ can readily be
obtained by numerically integrating the preceding expressions for
$g(q)$, $h(q)$, $g_{\fivesl ps}(q)$, and $j(q)$.  However, the scale of
$\alpha_S(q)$ still has to be set.  We turn to
this in the next section.

\newsec{Coupling Constant and Scale Determination}%
In this section we continue with the application of the Lepage-Mackenzie
prescription to determine the $\Lambda$-value of the coupling and
the scale~$q^*$ at which it is evaluated.
Although it is in principle a higher order correction in $\alpha_S$,
a large source of error in the renormalization of the matrix
element determining $f_B$ is which value of $\alpha_S$ to use.
The Lepage-Mackenzie prescription for fixing the value of the coupling
is to extract it from
a non-perturbative calculation of the
\hbox{1 $\times$ 1} Wilson loop ({\it i.e.,} the expectation value
of the plaquette, $U_{\fivesl plaq}$).  Once it is known at some scale
(alternatively, once its $\Lambda$ value is known), it can be run
to any other scale.  The formula which relates the Lepage-Mackenzie
perturbative coupling to the non-perturbatively determined (lattice
Monte Carlo) plaquette expectation value is,
\eqn\jq{
-\ln\vev{ \third {\rm Tr} U_{\fivesl plaq} }
= {4 \pi \over 3}  \alpha_V(3.41/a)
 \bigl[ 1- \alpha_V(3.41/a) (1.19 + 0.025 n_f) + O(\alpha_V^2) \bigr].
}
The coefficient of $n_f$ in Eq.~\jq\ is for dynamical Wilson
fermions with $r=1$.
The coefficient is $r$-dependent and is also different
for staggered fermions~\ref\urs{K.M. Bitar {\it et al,}
Florida State University preprint FSU-SCRI-93-110}.
However, in the application to the
quenched approximation, we set $n_f = 0$, and the value of the coefficient
does not enter our results.
The $\Lambda$ value for this coupling is denoted $\Lambda_V$.
The results of this determination are summarized in Table~1.
The values of $a^{-1}$ are those given by the charmonium
scale~\ref\charmonium{
A.~X.~El-Khadra, G.~Hockney, A.~S.~Kronfeld, and P.~B.~Mackenzie, Phys. Rev.
Lett. {\bf 69} (1992) 729.},
and the values of the plaquette
expectation value are taken from Ref.~\lm.

\topinsert
\vbox{
\def\tablerule{\noalign{\hrule}}
$$\vbox{\offinterlineskip\tabskip=0em
	\halign to 6in{\vrule #\tabskip=1em plus5em&
		\strut\hfil$#$\hfil&\vrule #&
      \hfil$#$\hfil&\vrule #&
      \hfil$#$\hfil&\vrule #&
      \hfil$#$\hfil&\vrule #&
      \hfil$#$\hfil&\tabskip=0em\vrule #\cr
		\tablerule
& \beta && a^{-1} {\rm (GeV)} && \vev{\third {\rm Tr}{U_{\fivesl plaq}}} &&
\alpha_V(3.41/a) && \Lambda_V
a &\cr
		\tablerule
		& 5.7 && 1.15(8)\phantom{1} && 0.549 && 0.1830 && 0.294 &\cr
		\tablerule
		& 5.9 && 1.78(9)\phantom{1} && 0.582 && 0.1595 && 0.198 &\cr
		\tablerule
		& 6.1 && 2.43(15) && 0.605 && 0.1450 && 0.144 &\cr
		\tablerule}}
$$
\centerline{Table 1. $a^{-1}$, $\alpha_V(3.41/a)$ and $\Lambda_V$
obtained from Monte}
\centerline{Carlo calculations of the charmonium spectrum~\charmonium\ and}
\centerline{the plaquette expectation value~\lm\ at various $\beta$.}}
\endinsert

It remains to fix the scale $q^*$, at which $\alpha_V(q)$ is evaluated,
Lepage and Mackenzie propose to do that by calculating the expectation value of
$\ln(q^2)$ in the one-loop perturbative lattice correction.  In equations,
\eqn\qstar{
\eqalign{
Y  & \equiv \int d^4 q ~\tilde g(q) \cr
\vev{\ln(qa)^2} & \equiv { \int d^4 q ~\tilde g(q) \ln(q^2a^2) \over Y } \cr
q^*a & \equiv \exp[\vev{\ln(qa)^2}/2] \cr
}}
Similar expressions for $Y_{\fivesl ps}$ and
$Y_{\fivesl ps}\vev{\ln{(qa)^2}}$ are defined with~$\tilde g_{\fivesl ps}(q)$
replacing~$\tilde g(q)$.  The only ambiguity in finding $q^*$ in either
case is how to deal with the constants
which do not arise from four-dimensional lattice integrals, and some
momentum-space integrations involving the heavy quark field
which were three-dimensional.  Fortunately these terms are generally
small relative
to the lattice contributions.  We write all such constants $C$ as
\eqn\constants{
C=\int d^4q {C\over (2\pi)^4} \ ,
}
and similarly take the $q_0$ dependence of the three-dimensional integrals
to be constant.  The values of the quantities just defined
are tabulated in Table~2a and~2b for various $r$.
These quantities were evaluated
using the Monte Carlo integration routine
VEGAS~\ref\VEGAS{G. P. Lepage, J. Comp. Phys. {\bf 27}, 192 (1978).}.
Errors on $Y$ and $Y \vev{\ln(q^2 a^2)}$
are at most $\curlyO(1)$ in the last decimal place.

\midinsert
\vbox{
\def\tablerule{\noalign{\hrule}}
$$\vbox{\offinterlineskip\tabskip=0em
	\halign to 4in{\vrule #\tabskip=1em plus5em&
		\strut\hfil$#$\hfil&\vrule #&
      \hfil$#$\hfil&\vrule #&
      \hfil$#$\hfil&\vrule #&
      \hfil$#$\hfil&\tabskip=0em\vrule #\cr
		\tablerule
		& r && Y && Y \vev{\ln(q^2 a^2)} && q^* a &\cr
		\tablerule
		&1.00 && -13.93 && -21.76 && 2.18 &\cr
		\tablerule
		&0.75 && -12.76 && -21.12 && 2.29 &\cr
		\tablerule
		&0.50 && -10.74 && -18.70 && 2.39 &\cr
		\tablerule
		&0.25 && -6.58 && -11.02 && 2.31 &\cr
		\tablerule
		&0.00 && \phantom{-}2.21 && \phantom{-1}9.65 && 8.88 &\cr
		\tablerule}}
$$
\centerline{Table 2a. $Y$, $\vev{\ln(q^2)}$, and $q^*$ for the zero distance}
\centerline{ lattice representation of $J^0_5$ defined in Eq.~\jofive.}
$$\vbox{\offinterlineskip\tabskip=0em
	\halign to 4in{\vrule #\tabskip=1em plus5em&
		\strut\hfil$#$\hfil&\vrule #&
      \hfil$#$\hfil&\vrule #&
      \hfil$#$\hfil&\vrule #&
      \hfil$#$\hfil&\tabskip=0em\vrule #\cr
		\tablerule
	& r && Y_{\rm ps} && Y_{\rm ps} \vev{\ln(q^2 a^2)} && q^* a &\cr
		\tablerule
		&1.00 && -7.20 && -10.87 && 2.13 &\cr
		\tablerule
		&0.75 && -5.20 && \phantom{1}{-8.10} && 2.18 &\cr
		\tablerule
		&0.50 && -2.13 && \phantom{1}{-2.81} && 1.93 &\cr
		\tablerule
		&0.25 && \phantom{-}2.98 && \phantom{-1}7.70 && 3.64 &\cr
		\tablerule
		&0.00 &&  \phantom{-}8.65 && \phantom{-}20.44 && 3.26 &\cr
		\tablerule}}
$$
\centerline{Table 2b. $Y_{\rm ps}$, $\vev{\ln(q^2)}$,
and $q^*$ for the distance-one}
\centerline{ lattice representation of $J^0_5$ defined in Eq.~\spac.}
}
\endinsert
\newsec{Application of Results to Lattice-Continuum Matching}%
In this section, we illustrate the application of the results
of the foregoing analysis
to the lattice-to-continuum part of the matching
for a couple of cases of interest.

Now that we have $\Lambda_V$ and $\alpha_V(3.41a^{-1})$
we use the two-loop formula with zero quark flavors ($n_f=0$),
to obtain $\alpha_V(q^*)$,
\eqn\twoloop{
\alpha_V(q) = \big[\beta_0 \ln(q^2/\Lambda_V^2)
               + (\beta_1/\beta_0) \ln\ln(q^2/\Lambda_V^2))\big]^{-1}
}
where $\beta_0=11/(4\pi)$ and $\beta_1=102/(4\pi)^2$.
Setting $n_f=0$ is appropriate for
lattice calculations done in the quenched approximation.  The explicit
dependence on the value of $a^{-1}$ drops out of the ratio $q^*/\Lambda_V$.
Hence, the only way we have used results from lattice Monte Carlo
calculations so far is for
the expectation value of the plaquette.
In Eq.~\ZandZps, which gives the lattice-to-continuum matching, we
put $\mu=q^*$, thus $\ln \mu a$ becomes $\ln q^* a$.
Thus the Monte Carlo calculation of the plaquette, and the
perturbative calculations are sufficient by themselves to determine
the lattice-to-continuum matching.
These results are summarized in Tables~3a and~3b.

\midinsert
\vbox{
\def\tablerule{\noalign{\hrule}}
$$\vbox{\offinterlineskip\tabskip=0em
	\halign to 4in{\vrule #\tabskip=1em plus5em&
		\strut\hfil$#$\hfil&\vrule #&
      \hfil$#$\hfil&\vrule #&
      \hfil$#$\hfil&\vrule #&
      \hfil$#$\hfil&\tabskip=0em\vrule #\cr
		\tablerule
		& r && \beta=5.7 && \beta=5.9 && \beta =6.1 &\cr
		\tablerule
		&1.00&& 0.73 && 0.77 && 0.80 &\cr
		\tablerule
		&0.75&& 0.76 && 0.80 && 0.82 &\cr
		\tablerule
		&0.50&& 0.82 && 0.84 && 0.86 &\cr
		\tablerule
		&0.25&& 0.91 && 0.92 && 0.93 &\cr
		\tablerule
		&0.00&& 1.13 && 1.11 && 1.11 &\cr
		\tablerule}}
$$
\centerline{Table 3a. $\tilde Z$ for various values of $r$ and $\beta$.}
$$\vbox{\offinterlineskip\tabskip=0em
	\halign to 4in{\vrule #\tabskip=1em plus5em&
		\strut\hfil$#$\hfil&\vrule #&
      \hfil$#$\hfil&\vrule #&
      \hfil$#$\hfil&\vrule #&
      \hfil$#$\hfil&\tabskip=0em\vrule #\cr
		\tablerule
		& r && \beta=5.7 && \beta=5.9 && \beta =6.1 &\cr
		\tablerule
		&1.00&& 0.88 && 0.90 && 0.91 &\cr
		\tablerule
		&0.75&& 0.93 && 0.94 && 0.95 &\cr
		\tablerule
		&0.50&& 1.00 && 1.00 && 1.00 &\cr
		\tablerule
		&0.25&& 1.13 && 1.11 && 1.10 &\cr
		\tablerule
		&0.00&& 1.24 && 1.21 && 1.19 &\cr
		\tablerule}}
$$
\centerline{Table 3b. $\tilde Z_{ps}$ for various values of $r$ and $\beta$.}
}
\endinsert

It still remains to multiply $\tilde Z$ and $\tilde Z_{\rm ps}$ by
the additional factors
summarized at the beginning of Section~3, then match the continuum
heavy quark effective theory at the lattice scale to the full
continuum theory at $m_b^*$.
We leave this for the conclusions, to which we now turn.

\newsec{Conclusions}%
In the preceding section, we illustrated the application of our one-loop
results to obtain the factors $\tilde Z$ and $\tilde Z_{ps}$, for the
local and point-split heavy-light bilinears.
{}From these $\tilde Z$'s we will obtain the
factors one multiplies a lattice calculation
by in order to obtain a physical number.

First, we must multiply by the two-loop running from $q^*$ to
$m_b^*$ and the matching that takes us from the
continuum heavy quark effective theory to the
continuum full theory.  The continuum results are explained
in detail in Refs.~\nextToLeading\ and~\bigpaper.  It is in this step that
the determination of the scale $a^{-1}$ is finally used.  The resulting
factor, $Z_{\fivesl cont}$, is $q^*$- and $a$-dependent.
The result after multiplying by this factor is tabulated in
Tables~4a and~4b.  Table~4a corresponds to using the zero-distance
representation of the axial current given in Eq.~\jofive, and
Table~4b corresponds to using the distance-one representation given
in Eq.~\spac.  Now that perturbation theory has been reorganized to
include tadpole corrections to all orders, we expect our one-loop calculation
of the renormalization factor to be accurate to
about 7\%.
This estimate of the magnitude of the two-loop correction was
obtained simply by squaring
the largest one-loop correction ($0.27^2=0.07$) in Tables 4a and 4b.

\midinsert
\vbox{
\def\tablerule{\noalign{\hrule}}
$$\vbox{\offinterlineskip\tabskip=0em
	\halign to 4in{\vrule #\tabskip=1em plus5em&
		\strut\hfil$#$\hfil&\vrule #&
      \hfil$#$\hfil&\vrule #&
      \hfil$#$\hfil&\vrule #&
      \hfil$#$\hfil&\tabskip=0em\vrule #\cr
		\tablerule
		& r && \beta=5.7 && \beta=5.9 && \beta =6.1 &\cr
		\tablerule
		&1.00&& 0.73 && 0.74 && 0.75 &\cr
		\tablerule
		&0.75&& 0.76 && 0.77 && 0.77 &\cr
		\tablerule
		&0.50&& 0.81 && 0.81 && 0.81 &\cr
		\tablerule
		&0.25&& 0.91 && 0.88 && 0.87 &\cr
		\tablerule
		&0.00&& 1.02 && 0.98 && 0.96 &\cr
		\tablerule}}
$$
\centerline{Table 4a. $\tilde Z \times Z_{\rm cont}$ for various values of $r$
and $\beta$.}
$$\vbox{\offinterlineskip\tabskip=0em
	\halign to 4in{\vrule #\tabskip=1em plus5em&
		\strut\hfil$#$\hfil&\vrule #&
      \hfil$#$\hfil&\vrule #&
      \hfil$#$\hfil&\vrule #&
      \hfil$#$\hfil&\tabskip=0em\vrule #\cr
		\tablerule
		& r && \beta=5.7 && \beta=5.9 && \beta =6.1 &\cr
		\tablerule
		&1.00&& 0.89 && 0.87 && 0.86 &\cr
		\tablerule
		&0.75&& 0.94 && 0.91 && 0.89 &\cr
		\tablerule
		&0.50&& 1.01 && 0.97 && 0.95 &\cr
		\tablerule
		&0.25&& 1.08 && 1.03 && 1.00 &\cr
		\tablerule
		&0.00&& 1.20 && 1.13 && 1.09 &\cr
		\tablerule}}
$$
\centerline{Table 4b. $\tilde Z_{ps} \times Z_{\rm cont}$ for various
values of $r$ and $\beta$.}
}
\endinsert

Second, we must multiply by the factors discussed in Section~3.
We multiply $\tilde Z$ and $\tilde Z_{\fivesl ps}$ by
$1/(2\sqrt{r})$,
and
$1/(2u_0\sqrt{r})$,
respectively.  The value of $u_0$ is obtained from
Eq.~\uodef\ and Table~1.  We find $u_0=0.861$, $0.873,$  and~$0.882$ for
$\beta=5.7$, $5.9$, and $6.1$, respectively.

As an example, consider the distance-zero operator at $\beta=6.1$.  From
Table~4a, we find $\tilde Z \times Z_{\rm cont}$ to be 0.75.  We multiply
by $1/(2\sqrt{r})=1/2$ to obtain the final result, $0.37$.  It is worthwhile
to compare this with the widely used value of~$Z_A$ of~$0.8$~\ehtwo,
which does not benefit from tadpole improvement.  To make
a head-to-head comparison it is necessary to multiply $Z_A=0.8$ by
$\sqrt{2\kappac}$ which is $0.56$ at
$\beta=6.1$~\ref\littlepaper{A.~Duncan,
E.~Eichten, J.~Flynn, B.~Hill, and H.~Thacker,  hep-lat/9312046,
FERMILAB--CONF--93/377--T, to appear in Nucl. Phys. B (Proc. Suppl.).}.
The product
is $0.45$.  Consequently, tadpole improvement results in a reduction of
the physical value of $f_B$ by a factor of $0.37/0.45$, {\it i.e.}, a
reduction of~18\%.
The change is similar to what one would have obtained using
non-tadpole-improved perturbation theory with a larger, ``boosted'' value
of $\alpha$, such as $\alpha_{\bar{\rm MS}}$ evaluated
at the scale of the inverse lattice spacing.
The direction of the change reduces the expected magnitude of the
$1/m$ corrections. These corrections must be included to fit
heavy quark effective field theory calculations and calculations
that extrapolate to the $b$-quark mass using propagating fermions with
various moderately heavy masses.
For a discussion of this extrapolation, see Ref.~\extrapolate.
Note that in that work a boosted coupling is used in the determination of~$Z$.

As mentioned in the introduction, both staggered and naive fermion results
correspond to the $r=0$ case~\hhone.  The
normalization convention for the Wilson fermion field is such that
the coefficient of $\bar\psi\psi(n)$ in the action is unity.  This term
is not present when $r=0$, and it is necessary to choose a different
normalization.  If one normalizes the field so that in the naive continuum
limit $\bar\psi \dsl \psi$ has coefficient one, then the
factor of $1/(2\sqrt{r})$ does not need to be included.
The factor of 1/$u_0$ remains necessary in the point-split case.

We have applied the Lepage-Mackenzie tadpole improvement program
to the case of the heavy-light quark current necessary to determine $f_B$
{}from lattice calculations.  We considered both the zero-distance
and distance-one point-split lattice
representations of $J^0_5$.
Our analysis included a slight generalization of the Lepage-Mackenzie
prescription to the case~$r\ne1$.
The results can be applied
whether the light quark is treated as a Wilson, staggered, or naive fermion.

\acknowledgements

We would like to thank Claude Bernard, Peter Lepage, and Paul Mackenzie
for many useful discussions, and Urs Heller for pointing out an
error in an intermediate formula in an earlier version of this paper.
OFH was supported by a National Science and Engineering Research
Council of Canada Canada International Fellowship.
BRH was supported in part by the Department of Energy under Grant No.
DE--FG03--91ER 40662, Task~C.
\vfill\eject
\baselineskip=16.8pt
\listrefs
\bye